\documentclass[prstab,twocolumn,showpacs,preprintnumbers,amsmath,amssymb,superscriptaddress]{revtex4}
\usepackage{graphicx}
\usepackage{dcolumn}
\usepackage{bm}
\usepackage[dvips]{epsfig}
\usepackage{subfigure}
\newcommand{\half}{\mbox{${\textstyle \frac{1}{2}}$}}

\begin{document}
\title{Determination of target thickness and luminosity from beam energy losses}
\author{H.J.~Stein}\affiliation{Institut f\"ur Kernphysik, Forschungszentrum J\"ulich,
  D-52425 J\"ulich, Germany}
\author{M.~Hartmann}\email{M.Hartmann@fz-juelich.de}
\affiliation{Institut f\"ur Kernphysik, Forschungszentrum
J\"ulich, D-52425 J\"ulich, Germany}
\author{I.~Keshelashvili}\affiliation{Institut f\"ur Kernphysik, Forschungszentrum J\"ulich,
  D-52425 J\"ulich, Germany}
\affiliation{High Energy Physics Institute, Tbilisi State
University, 0186 Tbilisi, Georgia}
\author{Y.~Maeda}\affiliation{Research Center for Nuclear Physics, Osaka
  University, Ibaraki, Osaka 567-0047, Japan}
\author{C.~Wilkin}\affiliation{Physics and Astronomy Department, UCL, London, WC1E 6BT, United Kingdom}
\author{S.~Dymov}\affiliation{Laboratory of Nuclear Problems, Joint Institute for
  Nuclear Research, RU-141980 Dubna, Russia}
\author{A.~Kacharava}\affiliation{Institut f\"ur Kernphysik, Forschungszentrum J\"ulich,
  D-52425 J\"ulich, Germany}
\affiliation{High Energy Physics Institute, Tbilisi State
University, 0186 Tbilisi, Georgia}
\author{A~Khoukaz}\affiliation{Institut f\"ur Kernphysik, Universit\"at M\"unster,
  D-48149 M\"unster, Germany}
\author{B.~Lorentz}\affiliation{Institut f\"ur Kernphysik, Forschungszentrum J\"ulich,
  D-52425 J\"ulich, Germany}
\author{R.~Maier}\affiliation{Institut f\"ur Kernphysik, Forschungszentrum J\"ulich,
  D-52425 J\"ulich, Germany}
\author{T.~Mersmann}\affiliation{Institut f\"ur Kernphysik, Universit\"at M\"unster,
  D-48149 M\"unster, Germany}
\author{S.~Mikirtychiants}\affiliation{Institut f\"ur Kernphysik, Forschungszentrum J\"ulich,
  D-52425 J\"ulich, Germany}
\affiliation{High Energy Physics Department, St.~Petersburg
Nuclear Physics Institute, RU-188350
  Gatchina, Russia}
\author{D.~Prasuhn}\affiliation{Institut f\"ur Kernphysik, Forschungszentrum J\"ulich,
  D-52425 J\"ulich, Germany}
\author{R.~Stassen}\affiliation{Institut f\"ur Kernphysik, Forschungszentrum J\"ulich,
  D-52425 J\"ulich, Germany}
\author{H.~Stockhorst}\affiliation{Institut f\"ur Kernphysik, Forschungszentrum J\"ulich,
  D-52425 J\"ulich, Germany}
\author{H.~Str\"{o}her}\affiliation{Institut f\"ur Kernphysik, Forschungszentrum J\"ulich,
  D-52425 J\"ulich, Germany}
\author{Yu.\,Valdau}\affiliation{Institut f\"ur Kernphysik, Forschungszentrum J\"ulich,
  D-52425 J\"ulich, Germany}
\affiliation{High Energy Physics Department, St.~Petersburg
Nuclear Physics Institute, RU-188350
  Gatchina, Russia}
\author{P.~W\"{u}stner}\affiliation{Zentralinstitut f\"ur Elektronik, Forschungszentrum
  J\"ulich, D-52425 J\"ulich, Germany}
%
\date{\today}

\begin{abstract}

The repeated passage of a coasting ion beam of a storage ring
through a thin target induces a shift in the revolution frequency
due to the energy loss in the target. Since the frequency shift is
proportional to the beam-target overlap, its measurement offers
the possibility of determining the target thickness and hence the
corresponding luminosity in an experiment. This effect has been
investigated with an internal proton beam of energy 2.65\,GeV at
the COSY-J\"ulich accelerator using the ANKE spectrometer and a
hydrogen cluster-jet target. Possible sources of error, especially
those arising from the influence of residual gas in the ring, were
carefully studied, resulting in a accuracy of better than 5\%.
The luminosity determined in this way was used, in conjunction
with measurements in the ANKE forward detector, to determine the
cross section for elastic proton-proton scattering. The result is
compared to published data as well as to the predictions of a
phase shift solution. The practicability and the limitations of
the energy-loss method are discussed.
\end{abstract}
%
%
\maketitle
%
%
\section{Introduction}
\label{sec:introduction}

In an ideal scattering experiment with an external beam, the
particles pass through a wide homogeneous target of known
thickness. If the fluxes of the incident and scattered particles
are measured, the absolute cross section of a reaction can be
determined. The situation is far more complicated for experiments
with an internal target at a storage ring where the target
thickness cannot be simply established through macroscopic
measurements. In such a case the overall normalization of the
cross section is not fixed though one can, for example, study an
angular dependence or measure the ratio of two cross sections. If
the value of one of these cross sections is known by independent
means, the ratio would allow the other to be determined. However,
there are often difficulties in finding a suitable calibration
reaction and so it is highly desirable to find an alternative way
to measure the effective target thickness inside a storage ring.

When a charged particle passes through matter it loses energy
through electromagnetic processes and this is also true inside a
storage ring where a coasting beam goes through a thin target a
very large number of times. The energy loss, which is proportional
to the target thickness, builds up steadily in time and causes a
shift in the frequency of revolution in the machine which can be
measured through the study of the Schottky spectra~\cite{Zapfe}.
Knowing the characteristics of the machine and, assuming that
other contributions to the energy loss outside the target are
negligible or can be corrected for, this allows the effective
target thickness to be deduced. It is the purpose of this article
to show how this procedure could be implemented at the COSY
storage ring of the Forschungszentrum J\"ulich.

The count rate $n$ of a detector system which selects a specific
reaction is given by
\begin{equation}
n = L \left(\frac{d\sigma}{d\Omega}\right) \Delta\Omega\,,
\label{eq:1}
\end{equation}
where $d\sigma/d\Omega$ is the cross section, $\Delta\Omega$ the
solid angle of the detector and $L$ the beam-target luminosity.
This is related to the effective target thickness $n_{T}$,
expressed as an areal density, through
\begin{equation}
L = n_B\,n_{T}\,, \label{eq:2}
\end{equation}
where $n_B$ is the particle current of the incident beam.

\begin{figure*}[t!]
\centering
\includegraphics[clip,width=0.6\textwidth]{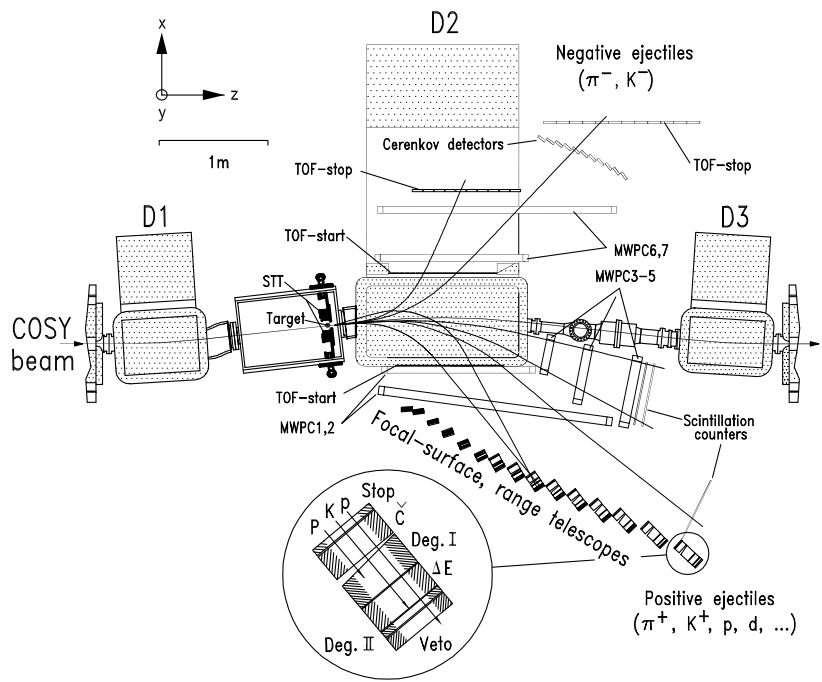}
\caption{Top view of the ANKE spectrometer and
detectors~\cite{Bar01,Har07}. The spectrometer contains three
dipole magnets D1, D2, and D3, which guide the circulating COSY
beam through a chicane. The central C-shaped spectrometer dipole
D2, placed downstream of the target, separates the reaction
products from the beam. The ANKE detection system, comprising
range telescopes, scintillation counters and multi-wire
proportional chambers, registers simultaneously negatively and
positively charged particles and measures their momenta. The
silicon tracking telescopes (STT) placed in the target chamber are
used to measure low energy recoils from the target.
\label{fig:anke}}
\end{figure*}

The luminosity, rather than the target thickness, is the primary
quantity that has to be known in order to evaluate a cross section
through Eq.~(\ref{eq:1}). The measurement of a calibration
reaction, such as proton-proton elastic scattering, leads directly
to a determination of the luminosity. In contrast, the energy-loss
technique described here yields directly an estimate of the
effective target thickness, but this can be converted into one of
luminosity through the measurement of the beam current, which can
be done to high accuracy using a beam current transformer.

Originally, the frequency-shift measurements were carried out and
analyzed at ANKE using only a few accelerator cycles over the
extended run of a specific experiment in order to get a rough
estimate of the available luminosity. However, a careful audit of
the various error sources has now been conducted to find out the
accuracy that can be achieved. Energy-loss measurements are
therefore now routinely carried out in conjunction with the
experimental data-taking.

\vspace{5mm} A brief presentation of the overall layout of the
ANKE spectrometer in the COSY ring is to be found in
Sec.~\ref{sec:ANKE} with the operation of COSY for this
investigation being described in Sec.~\ref{sec:machineoperation}.
The basic theory and formulae that relate the target thickness to
the change in revolution frequency are presented in
Sec.~\ref{sec:beamtargetinteraction}, where the modifications
caused by the growth in the beam emittance are also explained. The
application of the energy-loss method to the measurement of the
target thickness for typical target conditions when using a proton
beam with an energy of 2.65\,GeV is the object of
Sec.~\ref{sec:targetthickness}. A careful consideration is given
here to the different possible sources of error. These errors are
also the dominant ones for the luminosity discussed in
Sec.~\ref{sec:luminosity}. It is shown there that the relative
luminosity is already well determined through the use of monitor
counters so that the absolute luminosity given by the energy-loss
measurement needs only to be investigated for a sub-sample of
typical cycles. A comparison is made with the luminosity measured
through elastic proton-proton scattering at 2.65\,GeV, though this
is hampered by the limited data base existing at small angles. Our
summary and outlook for the future of the energy-loss technique
are offered in Sec.~\ref{sec:outlook}.

%
%
\section{COSY and the ANKE spectrometer}
\label{sec:ANKE}

COSY is a COoler SYnchrotron that is capable of accelerating and
storing protons or deuterons, polarized and unpolarized, for
momenta up to 3.7\,GeV/$c$, corresponding to an energy of 2.9\,GeV
for protons and 2.3\,GeV for deuterons~\cite{Mai97}.

The ANKE magnetic spectrometer~\cite{Bar01,Har07}, which is
located inside one of the straight sections of the
racetrack-shaped 183\,m long COSY ring, is a facility designed for
the study of a wide variety of hadronic reactions. The accelerator
beam hits the target placed in front of the main spectrometer
magnet D2, as shown in Fig.~\ref{fig:anke}. An assembly of various
detectors indicated in the figure allows, in combination with the
data-processing electronics, for the identification and
measurement of many diverse reactions. The method of determining
the luminosity from the beam energy loss in the target should be
applicable to the cases of the hydrogen and deuterium gas in
cluster-jet targets or storage cells that are routinely used at
ANKE. However, due to the short lifetime of the beam, the
technique is unlikely to be viable for the foil targets that are
sometimes used for nuclear studies.

%
%
\section{Machine Operation}
\label{sec:machineoperation}

\begin{figure}[htb]
\centering
\includegraphics[width=0.85\columnwidth,clip]{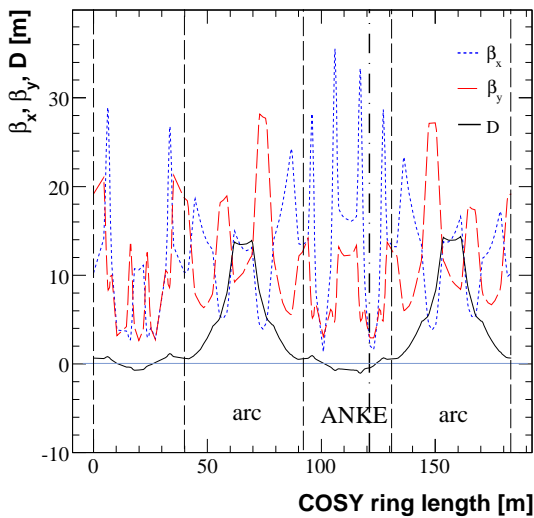}
\caption{(Color online) Ion optical functions around the COSY ring
as adjusted for experiments at ANKE. Here $\beta_x$ (dotted line),
$\beta_y$ (dashed line) are the horizontal and vertical beta
functions and $D$ (solid line) is the dispersion. \label{beta}}
\end{figure}

We discuss in detail the operational conditions of the 2004 beam
time where $\phi$-meson production in the $pp\to{}pp\phi$ reaction
was studied~\cite{Har06}. The proton beam with an energy of
2.650\,GeV was incident on a hydrogen cluster-jet target with a
diameter of 7\,mm~\cite{Mue04}. In order to accelerate the proton
beam from the injection energy of $T=45$\,MeV, a special procedure
is used at COSY which avoids the crossing of the critical
transition energy \mbox{$T_{\text{tr}}=mc^2
(\gamma_{\text{tr}}-1)$}~\cite{Mai97}.
For this purpose, a lattice setting that has a transition energy of about 
1\,GeV is used at injection. During the acceleration the ion optics in 
the arcs is manipulated such that the transition energy is dynamically 
shifted upward. After the
requested energy is reached, the acceleration (\emph{rf}) cavity
is switched off and the ion optics manipulated again such that the
dispersion $D$ in both straight sections vanishes. The transition
energy is then about 2.3\,GeV, \emph{i.e.}\ the experiment used a
coasting beam above the transition. Furthermore, the optics is
slightly adjusted to place the working point $(Q_x,Q_y)$ in the
resonance-free region of the machine between 3.60 and 3.66. This
guarantees that beam losses due machine resonances are avoided.
The resulting optical functions $\beta_x$, $\beta_y$, and
dispersion $D$ of the COSY ring, calculated within a linear optics
model, are shown in Fig.~\ref{beta}.

At the ANKE target position the parameters are $\beta_x = 2.4$\,m
and $\beta_y = 3.0$\,m. Orbit measurements have validated that the
dispersion is here within the range \mbox{$\pm\,0.5$}\,m. Since
$D\approx 0$ in this region, the ion beam does not move away from
the target when its energy decreases. The ion beam losses occur
dominantly in the arcs, where the machine acceptance is lower due
to the large dispersion of up to 15\,m. Experience has shown that,
depending upon the actual target thickness, experiments with the
cluster-jet target can be run with cycle times of 5-10 minutes
with little ion beam losses.

\begin{figure}[htb]
\centering
\includegraphics[width=0.85\columnwidth,clip]{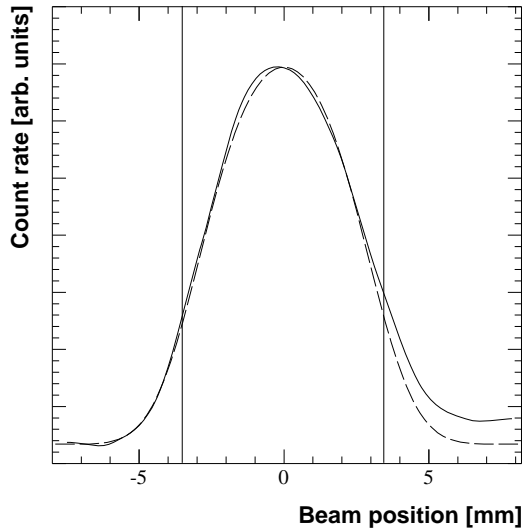}
\caption{Transverse beam-target overlap profile. The solid curve
shows the measured profile while the dashed line shows the
predicted one. The vertical lines represent the measured diameter
of the cluster-jet beam.\label{overlap}}
\end{figure}

The maximum of the beam-target interaction was found by steering
horizontally the proton beam continuously through the target and
identifying the highest count rate in the forward detector system
which was used as a monitor. The measured overlap profile shown in
Fig.~\ref{overlap} also contains information about the proton beam
size. The predicted profile was obtained by convoluting a
cylindrical cluster-jet beam of uniform density and 7\,mm
diameter~\cite{Mue04} with a Gaussian proton beam profile of width
$\sigma_x = 1.2$\,mm. Despite the idealized assumptions, the
measured profile is reasonably well reproduced. The maximum
overlap varies by less than 10\% for $\sigma_x$ in the range from
1.0 to 1.5\,mm.

The proton beam profile was independently investigated by scraping
the beam at the target position with a diaphragm oriented
perpendicular to the beam, which was moved through the beam. This
yielded a Gaussian beam profile with a total width $4\sigma\approx
5\,$mm~\cite{Leh04}. Later dedicated measurements have also
confirmed the typical size of the beam~\cite{Kirill}.

The beam-target interaction, \emph{i.e.}\ the effective target
thickness, might decrease during a machine cycle. This could arise
from emittance growth or the dispersion not being exactly zero and
would induce a slight nonlinear time dependence of the frequency
shift. Emittance growth and effective target thickness are
discussed in Secs.~\ref{sec:emittancegrowth} and
\ref{sec:targetthickness}.

\section{Beam-target interaction, energy loss and emittance growth}
\label{sec:beamtargetinteraction}

The fact that most ANKE experiments ran with a coasting beam
without cooling offered the possibility for using the energy loss
in the target as a direct and independent method for luminosity
calibration.

\subsection{Energy loss}
\label{sec:energyloss}

The energy loss $\delta T$ per single target traversal, divided by
the stopping power $dE/dx$ and the mass $m$ of the target atom,
yields the number $n_T$ of target atoms per unit area that
interact with the ion beam:
\begin{equation}
n_T = \frac{\delta T}{(dE/dx) m}\,\cdot \label{eq:3}
\end{equation}

Over a small time interval $\Delta t$, the beam makes $f_0\Delta
t$ traversals, where $f_0$ is the revolution frequency of the
machine. If the corresponding energy loss is $\Delta T$,
Eq.~(\ref{eq:3}) may be rewritten as:
\begin{equation}
n_T = \frac{\Delta T}{f_0(dE/dx) m\,\Delta t} \label{eq:4}
\end{equation}
or, in terms of the change in the beam momentum $p$, as
\begin{equation}
n_T = \left(\frac{1 + \gamma}{\gamma}\right) \frac{T_0\Delta
p}{f_0(dE/dx) mp_0\Delta t}\,, \label{eq:5}
\end{equation}
where $T_0$ and $p_0$ are the initial values of the beam energy
and momentum, and $\gamma = (1 - \beta^2)^{-1/2}$ is the Lorentz
factor.

In a closed orbit, the fractional change in the revolution
frequency is proportional to that in the momentum:
\begin{equation}
\frac{\Delta p}{p_0} = \frac{1}{\eta} \frac{\Delta f}{f_0}\,,
\label{eq:6}
\end{equation}
where $\eta$ is the so-called \emph{frequency-slip parameter}.

Putting these expressions together, we obtain
\begin{equation}
n_T = \left(\frac{1 + \gamma}{\gamma}\right)  \frac{1}{\eta}
\frac{1}{(dE/dx) m}\,\frac{T_0}{f_0^2}\,\frac{df}{dt}\,\cdot
\label{eq:7}
\end{equation}

In order to be able to deduce absolute values for the target
thickness on the basis of Eq.~(\ref{eq:7}), it is necessary to
determine $\eta$ with good accuracy. The revolution frequency
depends on the particle speed $\beta c$ and orbit length $C$
through $f = \beta c/C$ where, due to dispersion, $C$ is also a
function of the momentum. Defining $dC/C = \alpha\,dp/p$, we see
that
\begin{equation}
\frac{df}{f} = \left(\frac{1}{\gamma^2}- \alpha\right)
\frac{dp}{p}\,\cdot \label{eq:8}
\end{equation}

Here $\alpha$ is the so-called \emph{momentum compaction factor},
which is a constant for a given lattice setting. The point of
transition, where $df$ changes its sign, occurs when
$\alpha=1/\gamma^2$. Generally, $\alpha$ lies between 0 and 1, so
that $df$ is negative below and positive above transition. In
terms of $\alpha = 1/\gamma_{\text{tr}}^2$, the expression for
$\eta$ reads:
\begin{equation}
\eta = \frac{1}{\gamma^2} - \frac{1}{\gamma_{\text{tr}}^2}\,\cdot
\label{eq:9}
\end{equation}

The value of $\gamma$ is fixed by the beam momentum, which is
known with an accuracy on the order of 10$^{-3}$. The value of
$\gamma_{\text{tr}}$ is fixed for an individual setting of the
accelerator lattice used in the experiment. Near the transition
point $\eta$ is small and this is the principal restriction on the
applicability of the frequency-shift method.

An estimate for $\gamma_{\text{tr}}$ may be made using lattice
models but, to obtain more reliable values, a measurement of
$\alpha$ is indispensable. This is done by changing the magnetic
field $B$ in the bending magnets by a few parts per thousand and
using
\begin{equation}
\frac{\Delta f}{f} = \alpha\, \frac{\Delta B}{B}\,\cdot
\label{eq:10}
\end{equation}

\subsection{Emittance growth}
\label{sec:emittancegrowth}

In addition to energy loss, the beam also experiences emittance
growth through the multiple small angle Coulomb scattering in the
target. At each target traversal the emittance of the ion beam
increases slightly in both directions and, as a consequence, the
beam-target overlap may be reduced. As discussed in
Sec.~\ref{sec:machineoperation}, both $D$ and $D^{\prime}$ are
practically zero in the ANKE region. In this case, the rate of
emittance $\epsilon$ growth is given by~\cite{Hin89}:
\begin{equation}
\frac{d \epsilon}{d t} = \half f_0 \beta_T
\theta_{\text{rms}}^2\:, \label{eq:13}
\end{equation}
where $\beta_T$ represents the value of the beta function at the
position of the target, and $\theta_{\text{rms}}$ the projected
rms scattering angle for a single target traversal. The $1/2$
factor comes from integrating over the phases of the particle
motion in the ion beam.

The value of $\theta_{\text{rms}}$ can be estimated from
\begin{equation}
\theta_{\text{rms}} = Z \frac{14.1\,\textrm{MeV}}{\beta c p}
\sqrt{\frac{x}{X_0}}\:, \label{eq:12}
\end{equation}
where $Z$ is the charge number of the incident particle and
$x/X_0$ the target thickness in units of the radiation length
$X_0$~\cite{Hin89}.

The final rms beam width $w_f$ after an emittance growth $\Delta
\epsilon$ is given by
\begin{equation}
w_f = \sqrt{w_i^2 + \beta_T\,\Delta\epsilon}\:. \label{eq:14}
\end{equation}

Under typical experimental conditions of a proton beam incident on
a cluster-jet target containing $n_T = 2 \times
10^{14}\textrm{cm}^{-2}$ hydrogen atoms, an initial horizontal
width of $w_{x,i} = 1.2$\,mm increases to only 1.36\,mm over a
10\,min period. This suggests that the beam-target overlap or
effective target thickness should be constant to within 5\% and
that the frequency shift should show a linear time dependence.
%
%
\section{Measurement of target thickness by energy loss}
\label{sec:targetthickness}

\begin{table*}[hbt]
\caption{Parameters relevant for the target thickness evaluation
at 2.650\,GeV} \label{parameters}
\begin{tabular}{ll}
\hline
Parameters&Values\\
\hline $f_0 =$ initial revolution frequency& 1.57695\,MHz\\
$\beta =v/c =$ particle speed based on $f_0$ and  $C_{\text{nom}}
=183.493$\,m (including ANKE chicane)& 0.9652 \\
$\gamma = (1 - \beta^2)^{-1/2}=$ Lorentz factor&3.824\\
$p_0 = \beta \gamma mc =$ beam momentum&3.463\,\textrm{GeV/}c\\
$T_0 = (\gamma -1)mc^2 =$ beam kinetic energy
&2.650\,\textrm{GeV}\\
$\alpha =$ momentum compaction factor & $0.183\pm0.003$\\
$\eta =$ frequency-slip parameter evaluated from the measured value of $\alpha$ &$-0.115\pm 0.003$\\
$dE/dx =$ stopping power of protons in hydrogen gas &$4.108\,$MeV\,cm$^2$\,g$^{-1}$~\cite{NIS} \\
\hline
\end{tabular}
\vspace{0.5 cm}
\end{table*}

The parameters required for the estimation of the target thickness
for the experiment under consideration are given in
Table~\ref{parameters}. Here $\beta$, $\gamma$, $p_0$, and $T_0$
are determined by the measured revolution frequency and nominal
circumference of the accelerator and $dE/dx$ is evaluated from the
Bethe-Bloch formula as is done, e.g., in Ref.~\cite{NIS}. The
frequency shift $\Delta f$ is measured by analyzing the Schottky
noise of the coasting proton beam and the momentum compaction
factor $\alpha$, and hence the $\eta$-parameter, by studying the
effects of making small changes in the magnetic field.

The origin of the Schottky noise is the statistical distribution
of the particles in the beam. This gives rise to current
fluctuations which induce a voltage signal at a beam pick-up. The
Fourier analysis of the voltage signal, \emph{i.e.}\ of the random
current fluctuations, by a spectrum analyzer delivers frequency
distributions around the harmonics of the revolution frequency.
For this purpose we used the pick-up and the spectrum analyzer
(standard swept-type model HP 8753D) of the stochastic cooling
system of COSY~\cite{Sto}, which was operated at harmonic number
1000. During the experimental runs with a target, the Schottky
spectra around 1.577\,GHz were measured every minute over the
566\,s long cycle, thus giving ten sets of data per cycle. 
The frequency span was 600\,kHz and the resolution 1\,kHz.
The sweep time of the analyzer was set to 6\,s so that, to a good
approximation, instantaneous spectra were measured, which were
then directly transferred to the central data acquisition of ANKE
for later evaluation.

The spectrum analyzer measures primarily the Schottky noise
current, which is proportional to the square root of the number
$N$ of particles in the ring. The amplitudes of the measured
distributions are therefore squared to give the Schottky power
spectra, which are representative of the momentum
distribution~\cite{Bou87}. The centroids of these power spectra
yield the frequency shifts needed for the calculation of the mean
energy losses. It must be emphasized here that, by definition, the
Bethe-Bloch $dE/dx$ refers to the mean energy loss.

\begin{figure}[t!]
\centering
\includegraphics[width=0.85\columnwidth,clip]{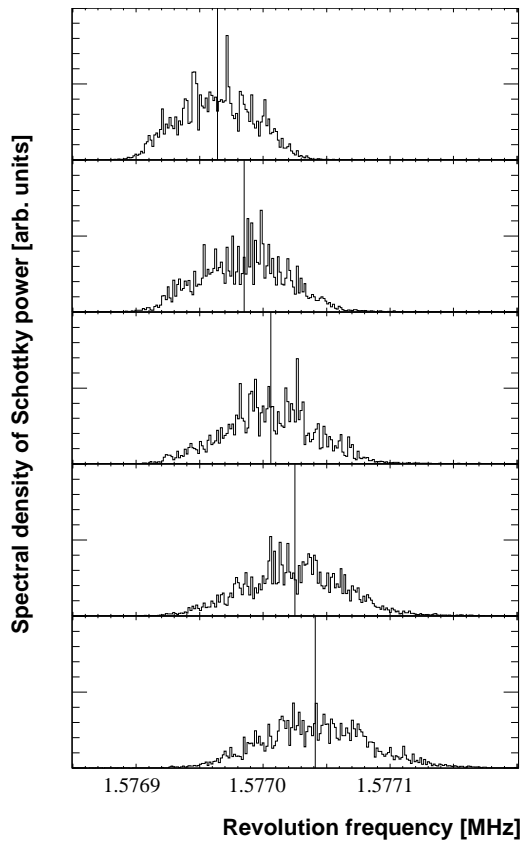}
\caption{Schottky power spectra obtained during one ten-minute
cycle and scaled to harmonic number 1. Although the data were
recorded every minute, for ease of presentation, only the results
from the even minutes are shown, starting from top to bottom. Each
spectrum is a true representation of the momentum distribution,
and the shift over the cycle is a measure for the energy loss. The
mean frequencies resulting from the fits are indicated by the
vertical lines. Since these data were taken above the transition
energy, $\eta$ is negative and the frequency increases through the
cycle.} \label{fig:schottky}
\end{figure}

Figure~\ref{fig:schottky} shows a typical result for the Schottky
power spectra obtained during one of the ten minute cycles. Due to
the momentum spread of the coasting beam, the spectra have finite
widths. The overall frequency shift in the cycle, which is
comparable to the width, is positive because at 2.65\,GeV the
accelerator is working above the transition point. Even the final
spectrum in Fig.~\ref{fig:schottky} fits well into the
longitudinal acceptance and there is no sign of any cut on the
high frequency side. The background was estimated by excluding
data within $\pm3\sigma$ of the peak position. After subtracting
this from the original spectrum, the mean values of the frequency
distribution was evaluated numerically.

\begin{figure}[htb]
\centering
\includegraphics[width=0.85\columnwidth,clip]{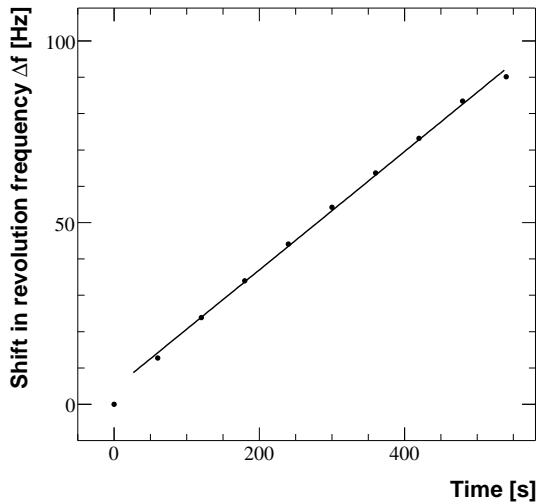}
\caption{Typical mean frequency shifts derived from the Schottky
power spectra of the type illustrated in Fig.~\ref{fig:schottky}
at ten equally spaced intervals of time. These results show a
linear increase with time with a slope $df/dt = (0.163\pm
0.003)\,$Hz/s. The first point was omitted from this fit since it
was taken too early in the cycle when the COSY magnets have not
reached their steady state after the acceleration.}
\label{frequency_variation}
\end{figure}

The time dependence of the mean revolution frequency shift $\Delta
f$ is shown for a typical cycle in Fig.~\ref{frequency_variation}.
It is well described by a linear function, which is consistent
with the assumption that the beam-target overlap changes little
over the cycle. This means that the emittance growth is negligible
and that there is no significant shift of the proton beam arising
from a possible residual dispersion. A linear fit over the
particular cycle considered here gives a slope of $df/dt =
(0.163\pm0.003)$\,Hz/s.

\begin{figure}[htb]
\centering
\includegraphics[width=0.85\columnwidth,clip]{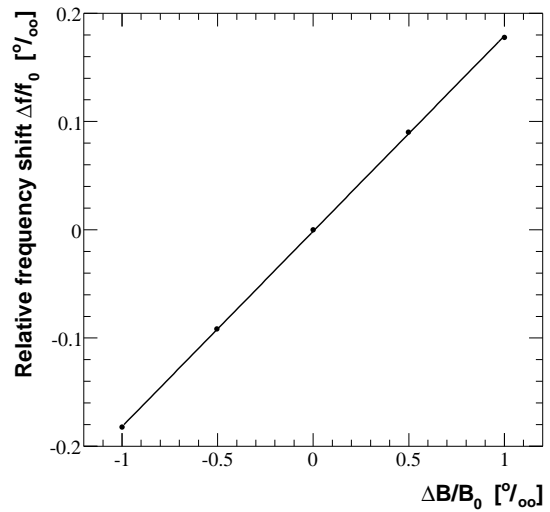}
\caption{Variation of the mean revolution frequency with the field
strength in the bending magnets in parts per thousand. The slope
of the fitted straight line yields the value of the momentum
compaction factor $\alpha$.} \label{eta}
\end{figure}

The value of the frequency-slip parameter $\eta$ was obtained by
measuring the momentum compaction factor $\alpha$ using separate
machine cycles without target. The shift of the mean revolution
frequency as a function of the $\Delta B/B_0$ change in the
bending magnets was investigated in the same way as for the energy
loss by determining the mean value of the frequency distributions.
Figure~\ref{eta} shows the five measured points for the relative
frequency shift $\Delta f/f_0$ as a function of $\Delta B/B_0$ in
the range from $-1.0$ to $+1.0$ per mille, in steps of 0.5 per
mille. The straight line fit, which is a good representation of
the data, leads to a value of the slope. These measurements were
carried out on three separate occasions during the course of the
four-week run and consistent values of the slope were obtained,
from which we deduced that $\alpha = 0.183 \pm 0.003$, and hence
$\eta = -0.115 \pm 0.003$.

Using Eq.~(\ref{eq:7}), a first approximation to the value of the
effective target thickness can now be given, assuming that the
measured frequency shift is dominantly caused by the target
itself. The result for the particular machine cycle, which is
typical for the whole run, is $n_T = 2.8 \times
10^{14}$\,cm$^{-2}$. This result contains, of course, a
contribution arising from the residual gas in the ring. The
systematic correction that is needed to take account of this is
discussed in the following section.
%
%
\subsection{Systematic correction for residual gas effects}
\label{sec:residualgas}

The contribution of the residual gas in the ring to the energy
loss was measured in some cycles with the target switched off. The
resulting frequency shift rate was $df/dt = (0.008 \pm
0.003)$\,Hz/s, which corresponds to a 5\% effect as compared to
that obtained with the target. The measurement was repeated a few
times during the four weeks of the experiment and the result was
reproducible to within errors. This is consistent with the
observation that the pressure in the ring was stable.

\begin{figure}[t]
\centering
\includegraphics[width=0.85\columnwidth,clip]{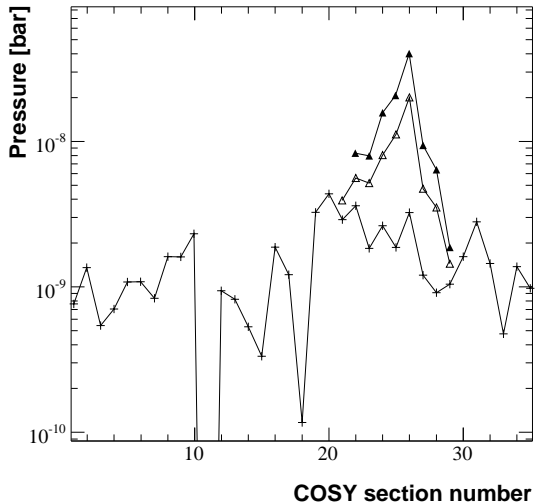}
\caption{Vacuum pressure profile along sections of the COSY ring;
ANKE is placed close to section number 26. Crosses show the
pressure profile for the situation when the (cluster-jet) target
beam is off, the increase illustrated by the open triangles is due
to the target-on effect, and the closed triangles that when the
COSY beam is allowed to interact with the target jet. The lines
are drawn to guide the eye.} \label{fig:pressure}
\end{figure}

However, as seen from Fig.~\ref{fig:pressure}, the gas pressure
rises in the vicinity of ANKE when the target is switched on. The
figure shows the vacuum pressure profile along the 183\,m long
ring for the three conditions (a) target off and no proton beam,
(b) target on and no proton beam, (c) target on and proton beam
incident on the target. A pressure bump with a maximum in the
target chamber region is spread over a region of about $\pm 5$\,m,
up- and downstream of the target position, which is in the
vicinity of section 26. The pressure in the target vacuum chamber
was $4 \times 10^{-9}$\,mbar with the target off, which is about
twice the average over the whole ring. With the target on this
pressure reached $2 \times 10^{-8}$\,mbar and further increased to
$4\times 10^{-8}$\,mbar when the proton beam interacted fully with
the target. The pressure rise is obviously caused by hydrogen gas
not being completely trapped in the gas catcher. The additional
pressure increase when the proton beam hits the target might be
attributed to hydrogen gas originating from the cluster-jet target
or from the chamber walls after hits by protons in the beam.

One critical question is how much of the energy loss is caused by
hydrogen atoms that are not localized in the target beam. This
effect was examined by steering the proton beam to positions to
the right and left of the target beam. The result was encouraging
since $df/dt$ increased only a little to a value of $(0.010 \pm
0.002)$\,Hz/s. We therefore take $df/dt = (0.012 \pm 0.004)$\,Hz/s
when the proton beam hits the target and the pressure is doubled.
As a cross check, the areal density of hydrogen atoms in a 10\,m
long path of hydrogen gas at the measured pressure of $4 \times
10^{-8}$\,mbar was calculated and compared to the areal density
found for the target. After making corrections for using the
pressure gauge with hydrogen rather than air, the areal density of
hydrogen atoms was found to be $4.8 \times 10^{12}$\,cm$^{-2}$.
Compared to the $n_T = 2.7 \times 10^{14}$\,cm$^{-2}$ initially
estimated, this is only a 2\% effect, which confirms the result
found from the frequency shift measurement.

It can be assumed that the contribution of hydrogen to the
residual gas is proportional to the target thickness.
Nevertheless, the uncertainty is large and the final modification
of $df/dt$ reads:
\begin{equation}
(df/dt)_{\text{corr}} = \zeta\left[(df/dt)_{\text{total}} -
(df/dt)_{\text{ring}}\right]\,, \label{final_correction}
\end{equation}
with $\zeta = 0.97\pm 0.02$.
%
%
\subsection{Uncertainties in the target thickness determination}
\label{sec:accuracythickness}

It is obvious from Eq.~(\ref{eq:7}) that the only significant
uncertainty in the determination of the overall target thickness
arises from the measurement of the frequency shift $\Delta f$.
This error is primarily instrumental in nature. The fit gives an 
uncertainty of $\pm 2\%$ for the total frequency shift. The systematic 
correction due to the residual hydrogen gas amounts to $\pm 2\%$. 
Depending on the variation of the target density during the whole
experiment, the relative error in the correction for the ring
vacuum was between 1.5 and 3\% and the machine parameter $\eta$
contributes a further $\pm 3\%$. These uncertainties, which are
summarized in Table~\ref{tab:errorDE}, stem from independent
measurements so that they can be added quadratically to give a
total of about 5\%. For the cycle under study, the corrected value
of the effective target thickness then becomes
\[n_T = (2.6 \pm 0.13) \times 10^{14}\textrm{cm}^2\,.\]

\begin{table}[htb]
\centering \caption{Individual contributions to the uncertainty in
the determination of the effective target thickness from the beam
energy losses. The total uncertainty has been obtained by adding
the individual elements quadratically.}
\bigskip
\label{tab:errorDE}
\begin{tabular}{lr}
\hline
Uncertainty & [\%]  \\
\hline
Frequency shift $\Delta f(t)$   & 2 \\
Residual gas (ring)   & (1.5 - 3) \\
Residual gas (target section) & 2 \\
Frequency-slip parameter ($\eta$)  & 3 \\
\hline
\textbf{Total}          & 5 \\
\hline
\end{tabular}
\end{table}

It should be noted that the ring gas effect in the present case was
only 5\% of that due to the target. If the target were much
thinner, the uncertainty arising from the residual gas
would dominate the total error.
%
%
\section{Luminosity deduced from the effective target thickness}
\label{sec:luminosity}

As seen from Eq.~(\ref{eq:2}), the luminosity can be deduced from
the effective target thickness by multiplying by the mean ion
particle current $n_B$ as determined in the same cycle.
%
%
\subsection{Particle current measurement}
\label{sec:beamflux}

The beam current $i_B=n_{B}e$ was measured by means of a high
precision beam current transformer (BCT) which was calibrated to
deliver a voltage signal of 100\,mV for a 1\,mA current. The BCT
signal was continuously recorded by the ANKE data acquisition
system \emph{via} an ADC. The accuracy of the BCT is specified to
be $10^{-4}$, though care has to be taken to avoid effects from
stray magnetic fields. The BCT was therefore mounted in a
field-free region of the ring and, in addition, was magnetically
shielded. It was calibrated with a current-carrying wire placed
between the beam tube and ferrite core of the BCT. Applying a
current from a high precision source in the range from $-10$ to
$+10$\,mA, the linearity and offset of the signal recorded in the
data acquisition system were $3 \times 10^{-4}$ and 0.2\,mV
(corresponding to 0.002\,mA), respectively. In comparison to the
uncertainty of the target thickness, the error in the measurement
of proton particle current is negligible since the beam current
was typically 10\,mA.
%
%
\subsection{Luminosity determination}
\label{sec:luminositydetermination}

\begin{figure}[htb]
\centering
\includegraphics[width=1.0\columnwidth,clip]{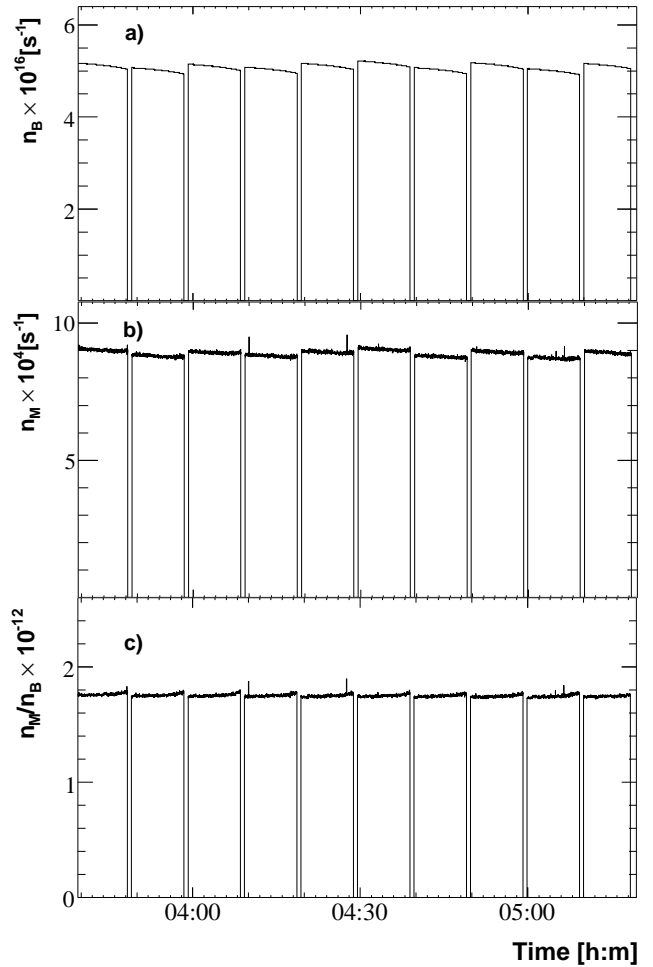}
\caption{(a) The BCT particle current $n_B$, (b) The monitor rate
$n_M$, and (c) The ratio $n_M/n_B$, for a sample of machine
cycles.} \label{fig:monitor}
\end{figure}

Figure~\ref{fig:monitor}(a) shows the proton particle current
$n_B$ for successive cycles. Within each cycle the current
decreases slightly with time due to beam losses from the
diminishing acceptance during the cycle which arise from the large
dispersion in the arcs. Since the initial beam current also varies
a little from cycle to cycle, the mean value $<n_B>$, and hence
the luminosity, has to be determined for each cycle. This yields
the mean or integrated luminosity over a certain period of time
which can then be compared directly with the results derived from
$pp$ elastic scattering or other calibration reaction.

Figure~\ref{fig:monitor}(b) illustrates the count rate $n_M$ of a
monitor for relative luminosity. For this purpose, the sum signal
of the start counters along the analyzing magnet D2 of
Fig.~\ref{fig:anke} has been selected. These counts originate
mainly from beam-target interactions, though there is some
background that does not come directly from the target.
Nevertheless, it is plausible to consider that the background rate
is also proportional to the proton beam intensity and target
density. That this is largely true is borne out by
Fig.~\ref{fig:monitor}(c), where the ratio of $n_M/n_B$ is
plotted. Except for a slight increase at the end of each cycle,
the ratio is constant within a cycle. This demonstrates that the
effective target thickness is constant, as already indicated by
the linear time dependence of the frequency shift. This behavior
was found to be true for all cycles in the experiment so that the
monitor count rate could be used as a good relative measure of the
luminosity over the whole experiment run. As a consequence, it is
sufficient to calibrate the monitor count rate by determining the
effective target density and mean ion particle current for only a
few representative cycles.

Since the measurement of the beam current with the BCT is accurate
to 0.1\%, the total uncertainty in the determination of the
luminosity \emph{via} the beam energy-loss method is 5\%, the same
as for the target thickness shown in Table~\ref{tab:errorDE}. The
values of the luminosity obtained during the experiment ranged
between 1.3 and $2.7\times
10^{31}\,\textrm{cm}^{-2}\textrm{s}^{-1}$.
%
%
\subsection{Comparison with proton-proton elastic scattering}
\label{sec:luminositypp}%

\begin{figure}[htb]
\centering
\includegraphics[width=0.85\columnwidth,clip]{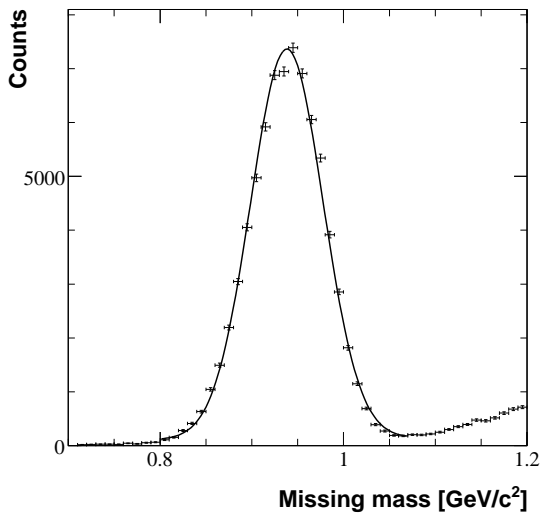}
\caption{Spectrum of missing masses measured for the $pp\to pX$
reaction at 2.65\,GeV showing a proton peak compared to a Gaussian
fit. This peak can be cleanly separated from the contributions
from pion production which start at 1.07\,GeV/$c^2$.}
  \label{fig:pMM}
\end{figure}

As an independent check on the energy-loss method, we have
measured the small angle elastic proton-proton differential cross
section. For this purpose the momentum of a forward-going proton
was determined using the ANKE forward detector, which covers
laboratory angles between about $4.5^\circ$ and $9.0^\circ$. The
large $pp$ elastic cross section, combined with the momentum
resolution of the forward detector, allows one to distinguish
easily elastically scattered protons from other events, as seen
from the missing-mass distribution shown in Fig.~\ref{fig:pMM}.

After making small background subtractions, as well as correcting
for efficiencies and acceptances, the number of detected $pp$
elastic scattering counts per solid angle, $dN_{pp}/d\Omega$, was
extracted as a function of the laboratory scattering angle. These
were converted into cross sections through Eq.~(\ref{eq:1}) using
the values of the luminosities deduced for each run using the
energy-loss technique. The individual contributions to the
systematic uncertainties in the cross sections are given in
Table~\ref{tab:errorpp}. If these are added quadratically, the
overall error is $\pm12\%$, which is twice as large as the error
in the luminosity determined by the beam-energy loss method.

\begin{table}[h]
\centering \caption{Systematic uncertainties in the measurement of
the cross section for $pp$ elastic scattering at $T_p =
2.65$\,GeV. The total error has been obtained by adding the
individual elements quadratically.}
\bigskip
\label{tab:errorpp}
\begin{tabular}{lr}
\hline
Uncertainty & [$\%$]  \\
\hline
Track reconstruction efficiency                & 5 \\
Acceptance correction                          & 8 \\
Momentum reconstruction                        & 1 \\
Data-taking efficiency                         & 5 \\
Background subtraction \hspace{2cm}            & 3 \\
Luminosity                                     & 5 \\
\hline
\textbf{Total}                                 & 12 \\
\hline
\end{tabular}
\end{table}

\begin{figure}[htb]
\centering
\includegraphics[width=0.85\columnwidth,clip]{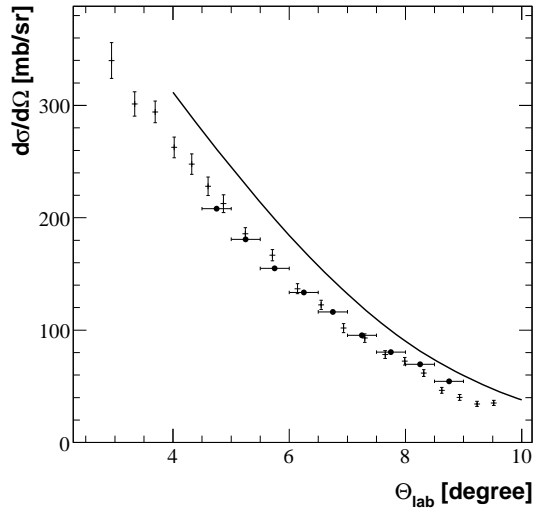}
\caption{Laboratory differential cross section for elastic
proton-proton scattering at 2.65\,GeV. Our points, shown by closed
circles with bin widths, have systematic uncertainties of $\pm
12$\%, as shown in Table~\ref{tab:errorpp}. The curve is the SP07
solution from the \textsc{SAID} analysis
group~\cite{SAID,SAID-update} and the crosses are experimental
data at 2.83\,GeV~\cite{Amb74}. }
  \label{fig:pp265}
\end{figure}

The values found for the proton-proton elastic differential cross
section at 2.65\,GeV are shown in Fig.~\ref{fig:pp265} together
with the current (SP07) solution obtained from the \textsc{SAID}
analysis group~\cite{SAID,SAID-update}. In general the
\textsc{SAID} program does not provide error predictions, but
these have been estimated by R.A.~Arndt~\cite{Arndt} to be on the
few percent level for our conditions.

The shape of the \textsc{SAID} curve is quite similar to that of
our data but these points lie about 20\% below the
predictions~\cite{SAID,SAID-update}. Such a discrepancy is larger
than the overall systematic uncertainty detailed in
Table~\ref{tab:errorpp}. It should also be stressed that the SP07
\textsc{SAID} solution also significantly overestimates the small
angle data of both Ambats \emph{et al.}~\cite{Amb74} at 2.83\,GeV
(shown in Fig.~\ref{fig:pp265}) and Fujii \emph{et
al.}~\cite{Fuj62} at 2.87\,GeV. It is therefore reassuring to note
the disclaimer in the recent \textsc{SAID} update, which states
that \emph{`our solution should be considered at best qualitative
between 2.5 and 3\,GeV'}~\cite{SAID-update}. This demonstrates
clearly the need for more good data in this region.
%
%
\section{Summary and outlook}
\label{sec:outlook}

We have shown that, under the specific experimental conditions
described here, the energy loss of a freely circulating (coasting)
ion beam interacting with a cluster-jet target can be used to
determine target thickness and beam-target luminosity. The method
is simple in principle and independent of the properties of
particle detectors which are involved in other techniques such as,
e.g., the comparison with elastic scattering. It relies on the
fact that the particles in a circulating beam pass through the
target more or less the same number of times so that they build up
the same energy shift. This is broadly true for the experiment
reported here, as can be seen from the fact that the Schottky
spectrum at the end of the cycle shown in Fig.~\ref{fig:schottky}
has a similar shape to that at the beginning.

Relative measurements of the luminosity are straightforward and
quick to perform during a run. The example given here involved the
ratio of a monitor rate $n_M$ and proton beam current $i_B$. Such
essentially instantaneous measurements have the advantage that
defective cycles with, e.g., a malfunction of the target, the ion
beam, or the detection system, can easily be removed from the data
analysis. The calibration of such relative measurements through
the energy-loss determination needs only to be done from time to
time and not for all runs.

The 5\% precision reported here for proton-proton collisions at
2.65\,GeV is mainly defined by the accuracy of the measured
frequency shifts. If the $pp$ elastic differential cross section
were known to say 5\%, it is seen from Table~\ref{tab:errorpp}
that the luminosity would only be evaluated using this information
at ANKE to about 12\%, which is much inferior to the energy-loss
method. However, the situation can be quite different at other
energies or for other targets.

\begin{figure}[htb]
\centering
\includegraphics[width=0.85\columnwidth,clip]{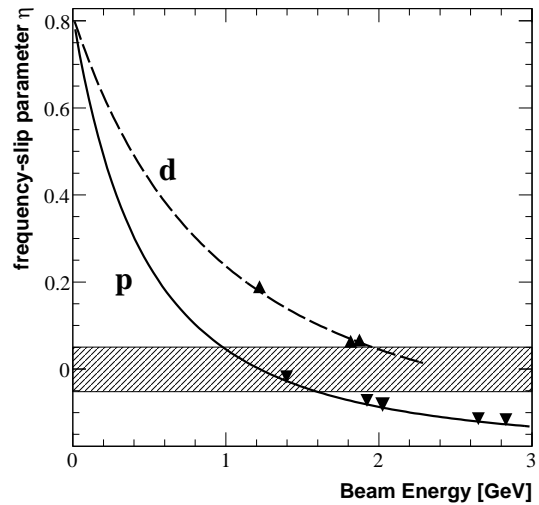}
\caption{Frequency-slip parameter $\eta$ as a function of the
energy of proton and deuteron beams. The experimental points are
the results of ANKE measurements during diverse beam runs. These
are compared with curves corresponding to the predictions of COSY
lattice calculations. The shaded area shows the region with
$|\eta|< 0.05$ where the error in the energy-loss technique can be
high.}
  \label{XXX}
\end{figure}

The relative error in the frequency-slip parameter $\eta$ of
Eq.~(\ref{eq:9}) becomes very large when $\gamma$ is in the region
of $\gamma_{\text{tr}}$. For the lattice setting normally applied
in ANKE experiments, where $\gamma_{\text{tr}} \approx 2.3$ and
the corresponding proton transition energy $T_{\text{tr}} \approx
1.2$\,GeV, the beam energy range from 1.0 to 1.6\,GeV is not well
suited for the energy-loss technique.

The application of the energy-loss technique to deuteron beams
and/or deuterium cluster-jet targets goes through identically. For
deuteron beams the method can be used over almost the whole of the
COSY energy range. This is illustrated clearly in Fig.~\ref{XXX},
which shows various measurements of the $\eta$ parameter for both
proton and deuteron beams compared with estimates from COSY
lattice calculations. The shaded area represents the region of
small $\eta$ where the method is of limited use.

The energy-loss method could be particularly valuable for
deuterons since, in such cases, there is often a lack of reliable
$pd$ elastic or quasi-elastic data~\cite{Mae06}. Furthermore, when
using small angle elastic $dp$ cross sections for normalization,
it has to be recognized that this varies exceedingly fast with
momentum transfer. As a consequence, even a small error in the
determination of the angle must be avoided or otherwise the
calibration can be seriously undermined~\cite{Timo2007}. Since the
energy loss is of electromagnetic origin, it could equally well be
used with beams of $\alpha$-particles or heavier ions.

The density of a cluster-jet target may be the ideal compromise
for implementing the energy-loss approach to luminosity studies.
Very thin foils are sometimes used as targets at
ANKE~\cite{Koptev} and the beam then dies too quickly for reliable
frequency shifts to be extracted. On the other hand, targets of
polarized gas in storage cells are very important for the future
physics program at ANKE~\cite{SPIN}. The overall target thickness
is less than that with the cluster jet so that the ring-gas will
provide a larger fraction of the energy loss. The ring-gas effects
will also be more important because of greater contamination of
the vacuum by the target. It is therefore clear that a detailed
analysis of the specific conditions is required to determine the
accuracy to be expected in a particular experiment.

%
%
\section*{Acknowledgments}
The detailed measurements reported here could only be carried out
with the active support of the COSY crew. We would like to thank
them and other members of the ANKE collaboration for their help.
Useful comments and information have come from I.~Lehmann.
R.A.~Arndt, I.~Starkovsky, and R.~Workman have
supplied updates on the \textsc{SAID} $pp$ data analysis and made
error estimates for our conditions. This work was supported in
part by the BMBF, DFG, Russian Academy of Sciences, and COSY FFE.
%
%

\end{document}